\documentclass[preprint,review,12pt]{elsarticle}

\usepackage{graphicx,float,multirow,hhline,fancyvrb,graphics,indentfirst,varioref,wrapfig,lettrine,subfigure}

\usepackage{soul}
\usepackage{amsmath,amssymb}
\usepackage[latin1]{inputenc}
\usepackage{enumerate}
\usepackage{epstopdf}
\usepackage{epsfig}
\usepackage{graphicx}
\usepackage{subfigure}
\usepackage{tikz}
\usepackage{pgfplots}

\usepackage{IEEEtrantools}
\usepackage{lineno}

\DeclareGraphicsExtensions{.pdf,.eps,.jpg,.png}

\usepackage[nodots]{numcompress}

\biboptions{sort&compress}

\journal{Renewable Energy}

\begin{document}

\begin{frontmatter}

\title{On the Aerodynamics of Multistage Co-Axial Vertical-Axis Wind Turbines}

\author[uoa]{Muhammad Saif Ullah Khalid\corref{mycorrespondingauthor}}

\cortext[mycorrespondingauthor]{Corresponding Author}
\ead{mkhalid1@ualberta.ca}

\author[uoc]{David Wood}

\author[uoa]{Arman Hemmati}

\address[uoa]{Department of Mechanical Engineering, University of Alberta, Edmonton T6G 1H9, AB, Canada}
\address[uoc]{Department of Mechanical and Manufacturing Engineering, University of Calgary, 2500 University Dr NW, Calgary, T2N 1N4, AB, Canada}

\begin{abstract}
This study explored the aerodynamics of a new multi-stage co-axial vertical-axis wind turbine based on bio-inspiration from natural swimming habit of fish. The turbine was formed from a conventional straight-bladed vertical axis turbine (VAWT)  with a small inner rotor, also of three blades.  The azimuthal  and radial locations of the inner rotor were varied. Using numerical simulations, performance of the proposed new design was evaluated over a range of tip-speed ratios. The preliminary results identified a $600\%$  increase in power output for multi-stage VAWTs at tip-speed rations $\mbox{TSR}=0-3$, and a substantial drop in power coefficient at $\mbox{TSR} > 0.3$. The wake dynamics analyses revealed that the increase was due to interactions between the blades of one rotor and the other. This reduced the unsteady separation from the outer rotor which produced most of the power. A detailed parametric study was also completed, which showed the implications of geometric and kinematic details on the performance of the proposed multistage VAWT.

\end{abstract}


\end{frontmatter}



\section{Introduction}
\label{sec1:Intro}

Energy is an essential commodity in defining our future prosperity. The scarcity of traditional resources and their limited sustainability, combined with their substantial contributions to climate change, have motivated extensive investment in research and development of alternative energy technologies. In this context, wind provides a great source of energy that is harnessed by different types of turbines. Vertical-axis wind turbines (VAWTs) are considered a great choice for energy harvesting in urban settings for small-scale power generation at lower wind speeds. Nevertheless, these turbines are also less efficient compared to horizontal-axis wind turbines (HAWTs) due to fluctuating aerodynamic loads, the consequent fatigue issues \cite{galinos2016vertical}, and continuous variations in the angle-of-attack ($\alpha$) for their blades with respect to the on-coming wind \cite{bazilevs2014fluid}. These features contribute to more complex aerodynamics of VAWTs. Moreover, VAWTs are categorized into Savonius, Darrieus, and H-rotor types. \citet{eriksson2008evaluation} presented a comparison of these turbines and argued that H-rotor type VAWTs offer more advantages in comparison to the other versions, including their simpler structures and no requirements of pitch regulators, yaw mechanisms, or gearboxes. However, these turbines also suffer from poor self-starting capabilities \cite{hill2009darrieus, asr2016study, sun2021rotation}. 

In order to increase the power output of VAWTs, various design modifications were proposed over the years. A popular strategy is to install VAWTs in parallel \cite{dabiri2011potential, zanforlin2016fluid, lam2017measurements, peng2020assessment, li2021experimental, hassanpour2021aerodynamic} or tandem configurations \cite{sahebzadeh2020towards, ni2021impacts}. \citet{dabiri2011potential} reported experimental investigations with VAWTs in counter-rotating arrangements and determined that closely-spaced turbines could attain higher power densities, even greater than those for HAWTs, by efficiently extracting energy from adjacent wakes. Through their two-dimensional numerical simulations, \citet{zanforlin2016fluid} found that the presence of another turbine in the vicinity of a VAWT in a side-by-side configuration modified the direction of on-coming wind in such a manner that its lateral velocity component favored the production of greater lift and torque. Similarly, \citet{lam2017measurements} conducted flow measurements in a wind tunnel for co-rotating and counter-rotating VAWTs and concluded that counter-rotating turbines helped maintain flow symmetry in the wake. In their case, staggered arrangements of turbines also showed small wake spreading rates and rapid wake recovery, which is advantageous in designing wind clusters and farms. Later, they also presented that power extraction capability of twin turbines was $8\% - 13\%$ greater than that of a solitary VAWT \cite{Lam2017b}. Recently, \citet{sahebzadeh2020towards} carried out three-dimensional simulations to find that an optimal spacing between two staggered rotors helped form a narrow region of high-speed flow in between them, which can increase the power coefficient of turbines installed in the downstream direction. Moreover, \citet{hassanpour2021aerodynamic} showed that in order to maximize the performance of staggered arrays, VAWTs should be positioned at the same height from the ground. Otherwise, differences in heights reduced their power production. 


Some researchers \cite{bhuyan2014investigations, ghosh2015computational, kumbernuss2012investigation} experimented with hybrid models of VAWTs. To improve the self-starting capability of VAWTs, \citet{bhuyan2014investigations} proposed a novel design by placing a Savonius rotors on top of a 3-bladed H-type rotor. This hybrid design was able to self start for all azimuthal angles by producing positive static torque. However, its power coefficient ($C_P$) depended on extent of the overlap of its two parts. The maximum value reached $0.34$, before dropping down due to further increase in overlap at tip-speed ratio ($\lambda$) of $2.29$ and Reynolds number ($\mbox{Re}$) of $1.92\times{10^5}$. The power output of each turbine was higher than the output of a single turbine. A similar approach was utilized by \citet{ghosh2015computational} to design a combined 3-bladed Darrieus-Savonius wind turbine for low-$\mbox{Re}$ flows in built environments. Also, \citet{kumbernuss2012investigation} employed two counter rotating Savonius turbines to develop a double stage VAWT with one stage directly above the other. The phase shift angle between the two stages in relation to wind speeds affected its performance significantly. More stages were added in this design of multi-stage Savonius turbine by \citet{saad2021performance}. It was reported that the maximum $C_P$ was $0.253$ and $0.261$ for two- and four-stage rotors, respectively, whereas a single-stage rotor obtained $C_P=0.223$. The greater advantage of the multistage systems was the reduction of oscillations in torque and thrust during rotational cycles, which was expected to mitigate fatigue and associate flow-induced noise, while improving the structural integrity of the system. This concept was also extended to H-rotor type VAWTs by \citet{didane2018performance, didane2019numerical}, where two 3-bladed rotors were installed together in co-axial contra-rotating settings. This configuration enabled tripling $C_P$ and torque compared to a single-stage turbine \cite{didane2018performance}. Next, they performed three-dimensional numerical parametric studies \cite{didane2019numerical} and found that keeping those two co-axial rotors close to each other increased their power output significantly. 

New designs of VAWTs can also be developed by installing rotors in series arrangements \cite{doppelter2011, torabi2016double, scungio2016wind, arpino2017cfd, arpino2018numerical, arpino2020numerical, su2020experimental, malael2018numerical, tahani2020unsteady}. The first attempt with this approach involved two co-axial coupled rotors installed in series \cite{doppelter2011}. Both outer and inner rotors contained 3 blades and the inner one rotated in the envelop of the outer rotor. Later, \citet{torabi2016double} performed two-dimensional simulations for the flow-induced rotation of coupled and uncoupled versions of these multi-stage VAWTs. Their results indicated that the presence of another rotor inside the primary one accelerated within smaller periods of time, which made it applicable to low-speed wind turbines. Moreover, \citet{malael2018numerical} and \citet{tahani2020unsteady} also employed similar configurations of two rotors in counter-rotating and co-rotating arrangements to explain their underlying flow dynamics. \citet{tahani2020unsteady} determined that the additional inner stage of the VAWT enhanced $C_m$ and $C_P$ by more than $300\%$. Perhaps, the most concerted efforts in this subject were carried out and reported by \citet{scungio2016wind} and \citet{arpino2017cfd, arpino2018numerical, arpino2020numerical}. First, \citet{scungio2016wind} showed that having three pairs of main and auxiliary blades instead of a conventional H-rotor type VAWT produced more dynamic torque for a large range of wind speeds. It also reduced the time taken to start the turbine from rest. \citet{arpino2018numerical} performed extensive wind tunnel testing and numerical simulations to demonstrate that their multi-stage turbine was able to harness sufficient energy for wind speeds lower than $4$ ${\mbox{m/sec}}$. Recently, \citet{su2020experimental} modified this design and found that connecting the auxiliary blades with leading edges of the main blades, and pitching them inwards, helped control power output of the turbine more effectively for varying wind speeds.          

These studies revealed great potential for improvements in currently available designs of VAWTs to enhance energy harvesting from natural winds in urban environments. Hybrid and multi-stage turbines remain the subject of extensive research. However, our understanding about these innovative systems is limited and more efforts are required to explain their underlying governing aerodynamic mechanisms. In this quest, we drew our inspiration from fish schooling phenomena to propose a new design of VAWTs with increased energy harvesting capabilities and better aerodynamic performance. It is well known \cite{Khalid2016a, gungor2020wake, gungor2021implications} that fish utilize specific configurations that enable significant advantages by harnessing more energy from the vortices generated by other fish in their vicinity. They tend to form various configurations to perform different social and hydrodynamic functions. These include circular arrangements, where individual members of schools position themselves in the form of co-axial circular loops with different radii, which resemble turbines from a two-dimensional perspective. \citet{pan2020computational} demonstrated that dense arrangements of the members in a fish school was more advantageous for their hydrodynamic performance. In the context of modern day technology, lessons learnt from natural schooling phenomena can also provide potentially effective solutions for various problems related to energy harvesting through tidal \cite{kinsey2012optimal} and wind turbines \cite{kinzel2012energy, brownstein2016performance}. Hence, we employed these ideas to computationally investigate dual-stage co-axial turbines in more details. Moreover, we explained the effects of various geometric and kinematic parameters on their performance and flow features. This manuscript is organized as follows. Section~\ref{sec:Nummeth} elucidated our computational methodology to perform simulations for single-stage and dual-stage VAWTs. Next, our findings on the performance of these VAWTs and their aerodynamic mechanisms were illustrated \label{sec:results}. Lastly, the summary and conclusions were presented in section~\ref{sec:concl}.

\section{Numerical Methodology}
\label{sec:Nummeth} 

The computational methodology used to perform simulations for flows over rotating turbines with prescribed angular velocities was described in this section. It includes discussions of the kinematics and details of the numerical setup, which was followed by detailed verification and validation studies.

\subsection{Geometry \& Kinematics}
\label{Geom} 
\begin{figure}[!ht]
\centering
\includegraphics[scale=0.5]{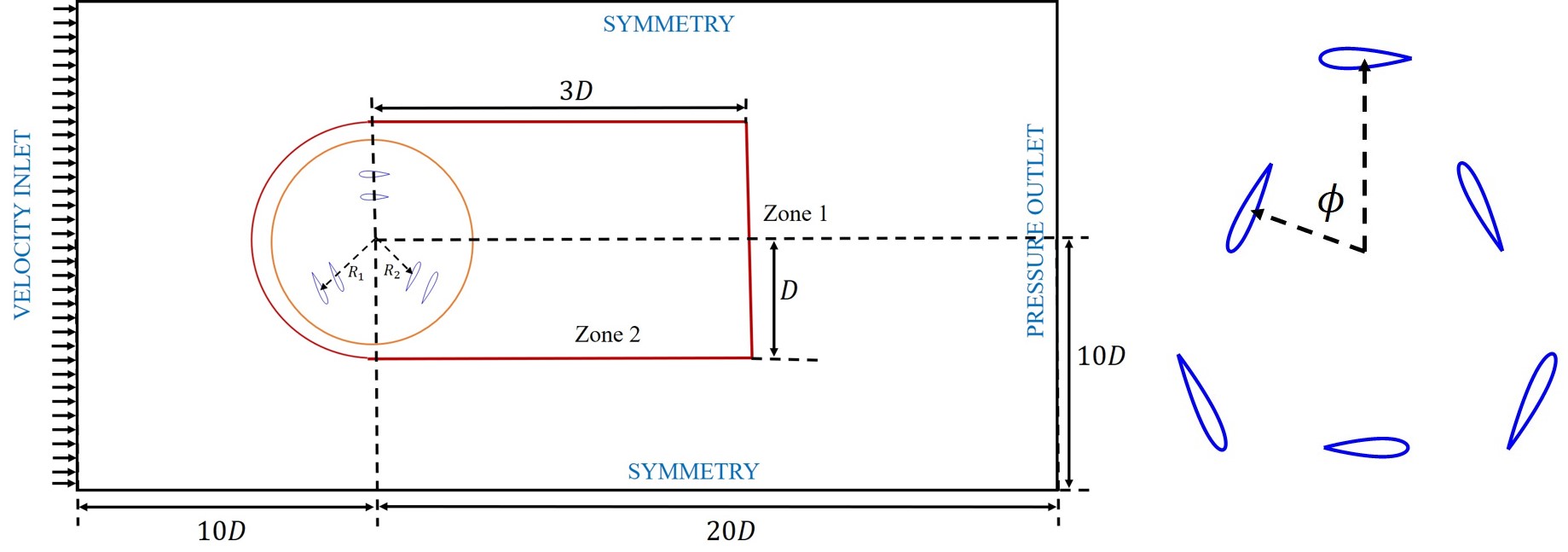}
\caption{A schematic representation of the computational domain with
details on boundary conditions} \label{fig:geometry}
\end{figure}

Single-stage and dual-stage configurations of H-type Darrieus vertical-axis wind turbines with a symmetric NACA0018 airfoil were used for this study, which represented the cross-section of each blade for this aerodynamic lift-based turbine. A schematic was provided in Fig. \ref{fig:geometry} with key kinematic and geometrical details outlined in Table~\ref{tab:geometry} along with the ranges of different governing parameters to determine the performance of VAWTs. 

\begin{table}[!ht]
\centering
\caption{Geometric and Flaw Parameters for VAWTs, where subscript "1" denotes the primary, outer rotor and "2" the inner, secondary one.} 
\centering
\begin{tabular}{c c}
\hline 
\hline
Parameters & Value \\
\hline
Airfoil section & NACA0018 \\
No. of Blades in single-stage turbines & $3$ \\
No. of Blades in dual-stage turbines & $6$ \\
Tip-speed ratio ($\lambda$) & $1.50-4.50$ \\
Free-stream velocity ($U_\infty$) & $7 \mbox{m/sec}$ \\
Chord length of blades in primary rotors ($c_1$) & $0.06\mbox{m}$ \\
Radius of primary rotors ($R_1$) & $0.5\mbox{m}$  \\
Ratios of radii for primary and secondary rotors & $R_1/R_2 = 0.85$ \& $0.92$ \\
Geometric ratios for blades in primary and secondary rotors & $c_1/D_1 = c_2/D_2 = 0.06$ \\
Angle between the blades of primary and secondary rotors ($\phi$) & $0^\circ$ - $90^\circ$\\
Reference area ($A$) & $1\mbox{m}^2$ \\
\hline
\hline
\end{tabular}
\label{tab:geometry}
\end{table}

The following parameters were commonly utilized in literature to describe the flow as well as the geometric and kinematic characteristics related to the wind turbines. First is the tip-speed ratio defined as: 

\begin{align}
\lambda={\frac{{\Omega}{R_1}}{U_\infty}}
\end{align}

\noindent where $R_1$ denotes the radius of the primary rotor of a turbine, $\Omega$ is the angular velocity, and $U_\infty$ represents the linear free-stream flow velocity. We tested the performance of VAWTs for $\lambda$ ranging from $1.50-4.50$, covering a broad range for wind turbine operation. For these kinematic conditions, \citet{Rezaeiha2017} showed that the results from $\mbox{2D}$ simulations approached those from $\mbox{2.5D}$ simulations. 

\subsection{Flow Solver}
\label{subsec:solver} 
The simulations were performed using ANSYS Fluent $2020 \mbox{R}2$ \cite{manual2020ansys}, a commercial finite volume based computational platform. This solver has gained a lot of popularity among researchers for wind turbine simulations \cite{Mohamed2014, Siddiqui2015, Ghasemian2015, Mohamed2016}. Incompressible unsteady Reynolds-averaged Navier-Stokes (URANS) and continuity equations were solved in Cartesian coordinates through the pressure-based solver: 

\begin{align}
\frac{\partial u_j}{\partial x_j} &=0 \label{eq:cont}
\end{align}

\begin{align}
\frac{\partial u_i}{\partial t} + \frac{\partial}{\partial
x_j}({u_i}{u_j}) &= -\frac{1}{\rho}\frac{\partial p}{\partial
x_i}+\nu \frac{{\partial^2}u_i}{{\partial x_j}{\partial x_j}}
\label{eq:NS}
\end{align}

\noindent where $x_j$ denotes the Cartesian coordinates, and $j=\{1,2\}$. Here, $u$ is the Cartesian velocity components, $\rho$ is the fluid density, $p$ is the pressure, and $\nu$ indicates the kinematic viscosity. Although the PISO (pressure implicit with splitting operators ) scheme is recommended for unsteady flows \cite{The2017}, it is usually advantageous when a large time-step (${\Delta}t$) is adopted to computationally march in time. Hence, Semi-Implicit Method for Pressure Linked Equation Consistent (SIMPLEC) algorithm was adopted for our present simulations to improve the computational efficiency.

The least square cell based technique was utilized for computation of gradient terms, second-order scheme for convective pressure terms, and second-order upwind technique for diffusion terms in the momentum equation (Eq. \ref{eq:NS}). Although third-order algorithms may also be used for the terms with Laplacian operator, those are computationally expensive. The advantage of upwind schemes is the provision of greater stability in numerical simulations. The unsteady term was approximated by the second-order implicit scheme.

First, the moving reference frame ($\mbox{MRF}$) approach was used to obtain steady-state flow features around the turbine, where the turbine did not actually rotate. The results from this analysis were then used as the initial condition to carry out unsteady simulations. This approach helped attain faster convergence of iterative solutions at each time-step. We performed unsteady simulations through the sliding mesh technique, which allowed physical rotation of the turbine without disturbing the original mesh.

\citet{rezaeiha2019accuracy} suggested that Shear Stress Transport (SST) turbulence models performed well in capturing the flow features for VAWTs and their results obtained through them closely matched with experiments. Therefore, SST- $k$- $\omega$ model was utilized to predict the turbulent flow features. Developed by \citet{menter1994two}, this model combines the robustness and accuracy of the $k$- $\omega$ model in near-wall regions with free-stream independence of the $k$- $\epsilon$ model in the far field. This blended formulation refines the standard $k$- $\omega$ model by modifying the definition of turbulent viscosity to incorporate the transport of turbulent shear stress. It enhances the accuracy and reliability of turbulence modeling for a wide range of flows with adverse pressure gradients.  

The convergence criterion for the iterative solution at each time-step was set to $10^{-4}$. Although we obtained convergence within $10-15$ iterations at each time-step, maximum allowable number of iterations were $50$. All the simulations were completed for $22$ revolutions, which provided the statistical quantities based on the data of last $5$ revolutions. As also discussed by \citet{Rezaeiha2017}, the steady-state solutions were achieved within $15-20$ revolutions of the turbine. 


\subsection{Computational Domain and Boundary Conditions}
An H-grid technique with a rectangular computational domain was used in this study, which is shown in Fig.~\ref{fig:geometry}. A uniform flow-velocity ($U_\infty=7\mbox{m/sec}$) was specified at the inlet boundary placed at a distance of $10D$ from the axis of the turbine. Gauge pressure was set as zero on the pressure outlet boundary that is $20D$ away from the rotational axis. Top and lower boundaries were set as symmetry boundaries, and each was at a distance of $10D$ from the turbine axis. We adjusted all domain boundaries following the recommendations of \citet{Rezaeiha2017}.

\begin{figure}[!ht]
\centering
\includegraphics[scale=0.2]{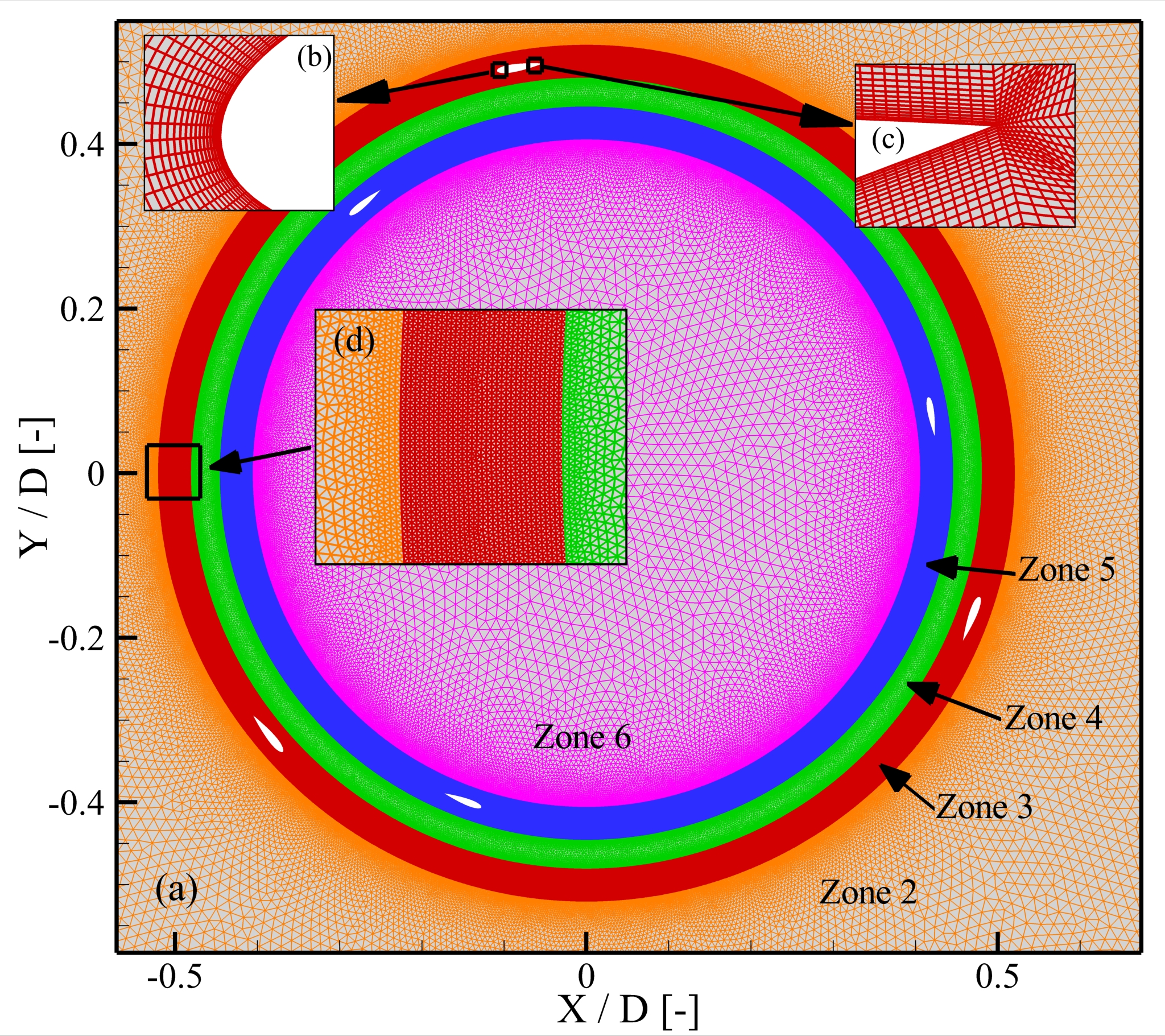}
\caption{A schematic representation of the computational domain with
details on boundary conditions} \label{fig:grid}
\end{figure}

To incorporate the moving reference frame and sliding mesh techniques, the computational domain was divided into five zones for dual-stage VAWTs. In Fig.~\ref{fig:geometry}, zone 1 and zone 2 are shown. Both these zones remained stationary and were connected by an interface between them, which enabled communication of flow information between neighboring domains through a non-conformal meshing algorithm. Zone 2 was used to capture details of the flow characteristics in the wake of the rotating turbine. Figure~\ref{fig:grid}a presents the meshing features inside the remaining zones 3-6. Here, zone 3 and zone 5 contained blades for the outer and inner stages of the turbine, respectively. These domains rotated around the central axis of the VAWT. However, domains represented by zones 4 and 6 remained stationary. Figures~\ref{fig:grid}b and \ref{fig:grid}c exhibited grid near the leading and trailing edges of the foils with ${y^+}=1$, respectively, which was sufficient to accurately capture boundary layer around these rotating structures. It is evident from Fig.~\ref{fig:grid}d that we controlled the mesh settings in order to slowly vary the mesh size and avoid large gradients. Numerical errors were avoided by keeping the same cell size around the interface boundary between the rotating and static domains. The current setup followed the recommendations of \citet{Rezaeiha2017}, whose study revealed that radius of a rotating domain had no significant implication on the aerodynamics of wind turbines in simulations. 




\subsection{Performance Parameters}
To measure the aerodynamic performance of wind turbines, nondimensional power and torque coefficients, denoted as $C_P$ and $C_T$, respectively, were computed for all cases, which were defined as: 

\begin{equation}
\begin{aligned}
C_T = \frac{T}{{q}{A}},\\
C_P = \frac{{T}{\Omega}}{{q}{U_\infty}{A}} 
\end{aligned}
\end{equation}

\noindent where $T$ denotes torque of the VAWT, $A$ is the swept area of the turbine, and $q={\rho}{{U^2}_\infty}/{2}$ represents the dynamic pressure. Swept area was a factor computed through multiplication of the turbine height ($1\mbox{m}$ for the current 2D cases) and its outer diameter. Using time-period ($\tau=1/f$) of one revolution, corresponding time-averaged coefficients were computed using the following relation:

\begin{align}
\bar{C}={\frac{1}{\tau}}{\int_t^{t+\tau} {C(t)} dt}
\end{align}

Steady-state power coefficient was calculated using the relation $\overline{C_P} = (\mbox{TSR})\overline{C_T}$. The statistical quantities were computed for the last $5$ revolutions of the VAWTs, where the variations in the time-averaged quantities were negligible \cite{Rezaeiha2017}. 

\subsection{Grid Independence Study}
\label{subsec:Grid} 
A detailed sensitivity study was completed to ensure grid convergence for the dual-stage VAWTs. The computational domain was composed of unstructured triangular cells in the fluid domain with $26$ layers of quadrilateral elements around each blade to resolve the boundary layer. We controlled the grid size by changing the maximum sizes of the grid in different zones while keeping $y^{+}$ value of the order of $1$. This parameter helped estimate the first cell height from the solid surface. Its accuracy is important in resolving the viscous sublayer in turbulent boundary layers. To carry out verification of the simulations results, three mesh sizes were considered for $\lambda=4.5$, the details of which are provided in Table~\ref{tab:gridconvrg}.

\begin{table}
\centering
\caption{Details of mesh in different zones for grid-independence study} 
\centering
\begin{tabular}{c c c c}
\hline 
\hline
Details of Grid               & G1 Coarse & G2 Medium & G3 Fine \\
\hline
Mesh Nodes on Each Blade       &     400   &   400    & 400 \\
Maximum Size in Rotating Zones &    0.002  &  0.00135 & 0.001 \\
Maximum Size in Zone 4         &    0.0065 &  0.0575  & 0.005 \\
Maximum Size in Zone 6         &    0.02   &  0.015   & 0.01 \\
Maximum Size in Zone 2         &    0.02   &  0.015   & 0.01 \\
Total Number of Cells          &    366164 &  600721  & 1015919 \\
\hline
\hline
\end{tabular}
\label{tab:gridconvrg}
\end{table}

\begin{figure}[!ht]
\centering
{\includegraphics[scale=0.22]{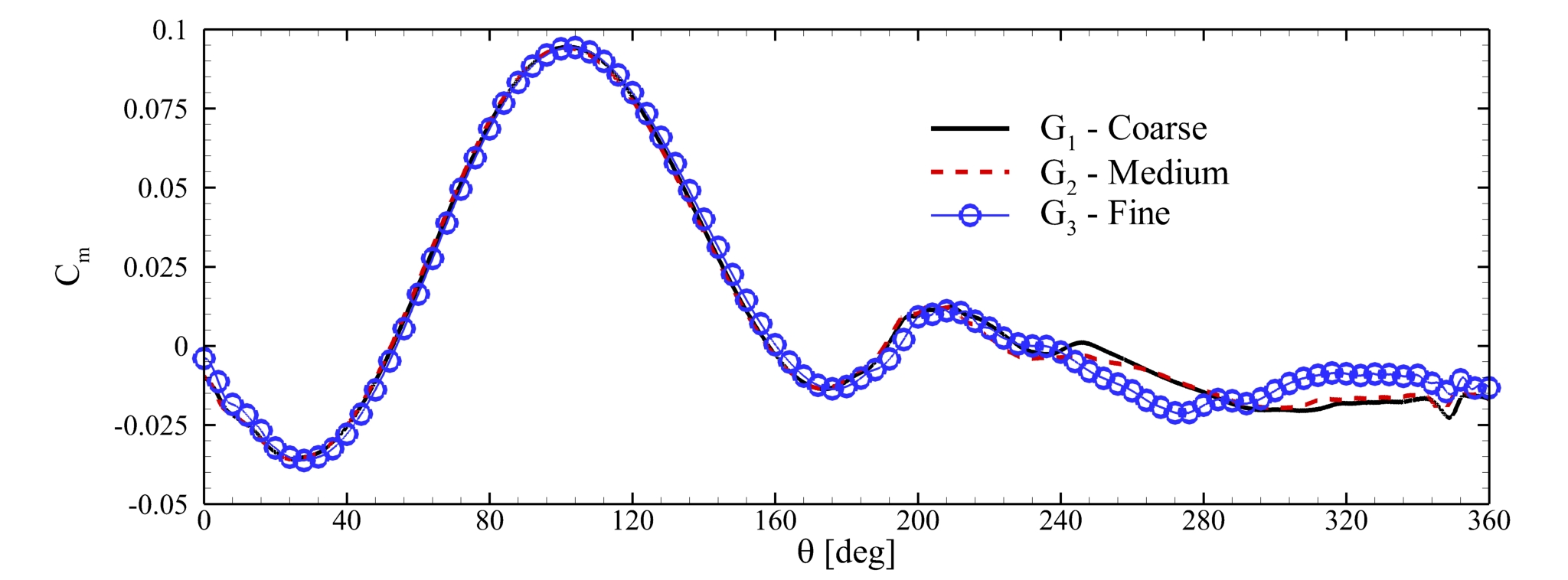}}
\caption{$C_m$ of a single blade for different grid configurations}
\label{fig:grid_indpndnc}
\end{figure}

Figure~\ref{fig:grid_indpndnc} presents the variations in the moment coefficient of one blade in a dual-stage rotor, undergoing rotation with $\lambda = 4.0$. It is apparent that all three meshes produced the same $C_m$ and small differences arose only for $ {240^\circ} < \theta < {350^\circ}$. Because the grid configuration $G_2$ matched more closely to $G_3$, we ran the remaining simulations with the mesh settings of $G_2$ with the change of azimuth angle $d\theta = 0.2^\circ$ at each time-step \cite{Rezaeiha2017, Rezaeiha2017b}.

\subsection{Validation}
\label{subsec:Validation}
\begin{figure}[!ht]
\centering
\includegraphics[scale=0.3]{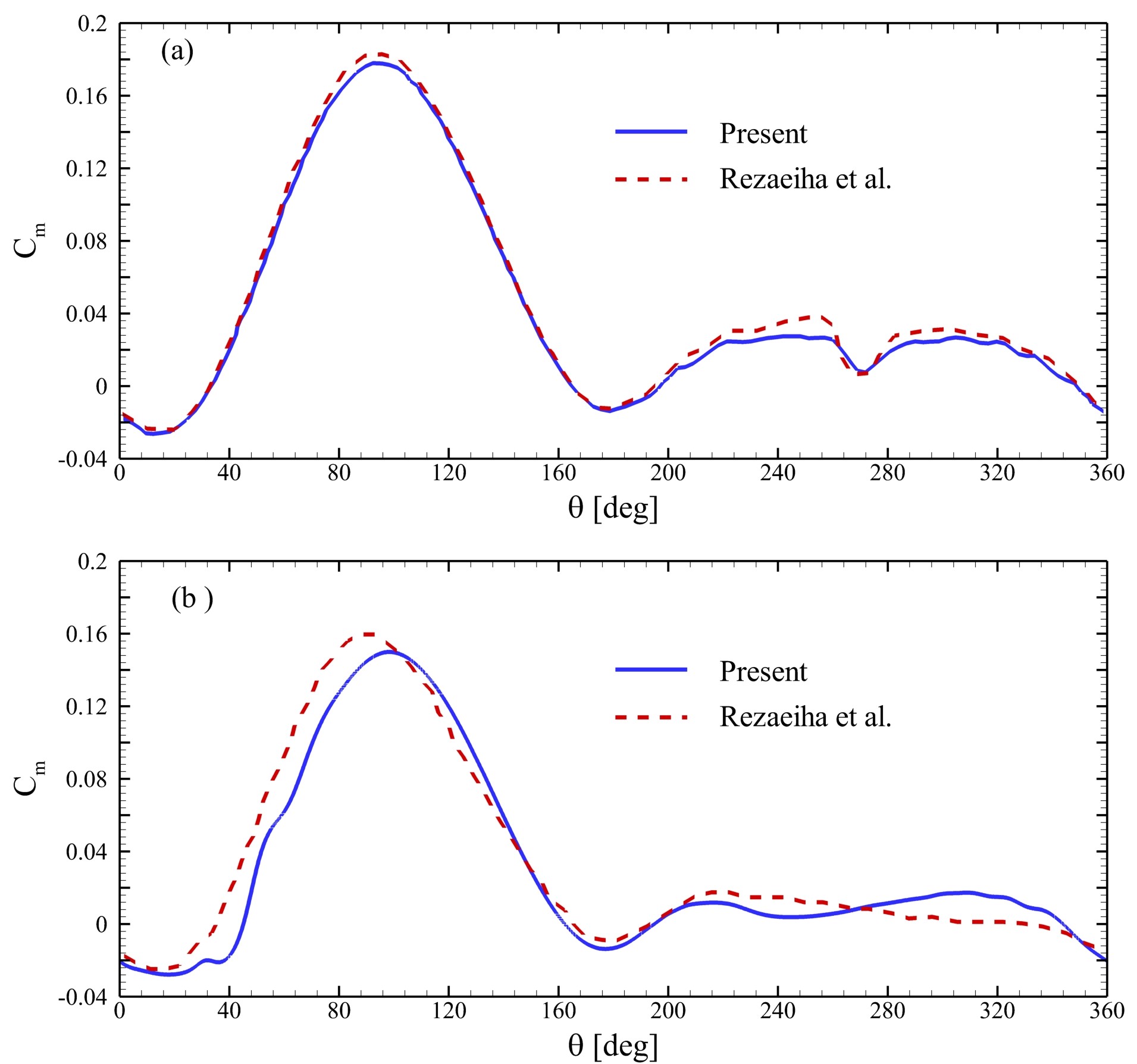}
\caption{Comparison of $C_m$ with those from numerical simulations of flows around a (a) 2-bladed VAWT \cite{Rezaeiha2017b} and (b) 3-bladed turbine \cite{Rezaeiha2017}}  
\label{fig:Validation_Cm}
\end{figure}

The simulations were validated by initially comparing their results with those reported in Ref. \cite{Rezaeiha2017} and Ref. \cite{Rezaeiha2017b}, for 2-bladed and 3-bladed single-stage turbines, respectively. Figure~\ref{fig:Validation_Cm}a exhibits $C_m$ of a single blade in a 2-bladed VAWT obtained by the present simulation settings and those of \citet{Rezaeiha2017}. It is evident that both profiles matched well for the complete rotation of the blade in Fig.~\ref{fig:Validation_Cm}. Additionally, $C_m$ followed similar trends for the 3-bladed VAWT with minimal differences in magnitude. Moreover, simulations for VAWTs used by \citet{Castelli2011} were repeated to compare the computational results with their experimental and numerical results. Table~\ref{tab:valid} shows that the current numerical values of $C_P$ were closer to the experimental measurements for even low ranges of tip-speed ratios.

\begin{table}
\centering
\caption{Comparison of the $\bar{C_p}$ from present results with those from literature}
\begin{tabular}{c c c c c}
\hline \hline
TSR & Exp. (Castelli et al. \cite{Castelli2011}) & CFD (Castelli et al. \cite{Castelli2011}) & Present \\
\hline
\hline
1.44 & 0.013 & 0.17 & 0.096 \\
1.68 & 0.044 & 0.251 & 0.138 \\
2.04 & 0.138 & 0.431 & 0.291 \\
\hline
\hline
\end{tabular}
\label{tab:valid}
\end{table}

\subsection{Effect of Turbulence Models}
\label{subsec:Validation}
\begin{figure}[!ht]
\centering
\includegraphics[scale=0.2]{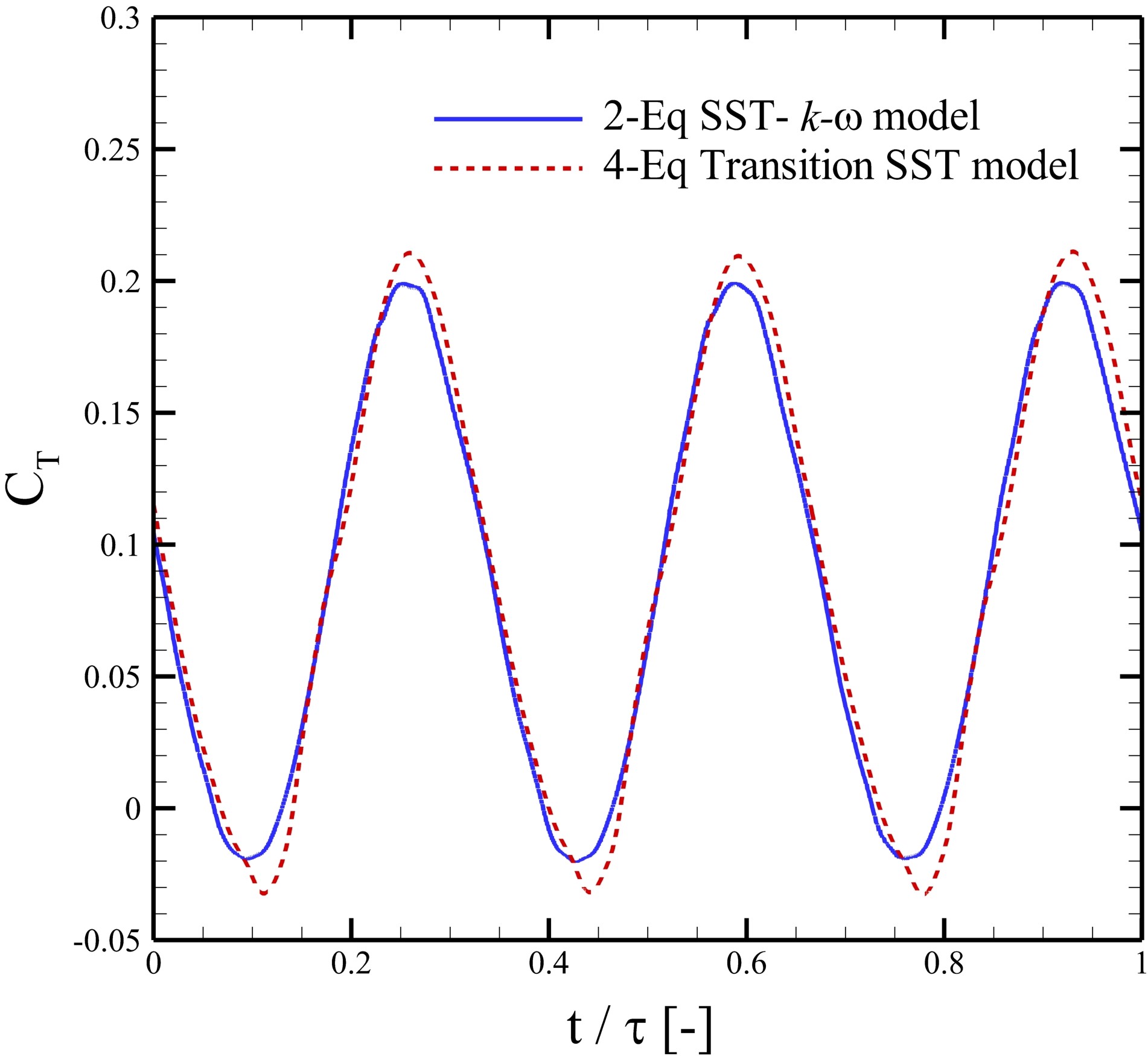}
\caption{Comparison of the torque coefficients ($C_T$) computed using two-equations SST- $k$- $\omega$ model and 4-equations transition SST model for flows over for a multi-stage VAWT with $\phi=0^\circ$ and ${R_2}/{R_1}=0.85$ at $\mbox{TSR}=3.0$}  
\label{fig:turb_cmprsn}
\end{figure}

In order to demonstrate the suitability of the turbulence model, a comparison of steady-state torque coefficients ($C_T$) for the whole turbine around its central axis computed through 2-equations SST- $k$-$\omega$ and 4-equations transition SST models was presented in Fig.~\ref{fig:turb_cmprsn}. Here, $\tau$ denoted the time-period for a complete revolution of the turbine. The plots clearly showed that the torque profiles remained largely unaffected by the choice of the turbulence model. The only minimal difference was observed at the time instants when $C_T$ attained its maximum and minimum values. Previously, \citet{rezaeiha2019accuracy} presented a detailed analysis of the effectiveness for difference turbulence models. They found that the employment of transition SST models captured the laminar-to-turbulence transition quite well. However, the results of all SST-based turbulence models matched closely with those obtained through experiments for a wide range of flow and kinematic governing parameters. Because the employment of 3-equations and 4-equations model are known to increase the computational cost by $14\%$ and $30\%$, respectively \cite{rezaeiha2019accuracy}, we chose the 2-equations SST- $k$-$\omega$ model to proceed with the remaining simulations. 

\section{Results \& Discussion}
\label{sec:results}
\begin{figure}[!ht]
\centering
\subfigure[]{\includegraphics[scale=0.1]{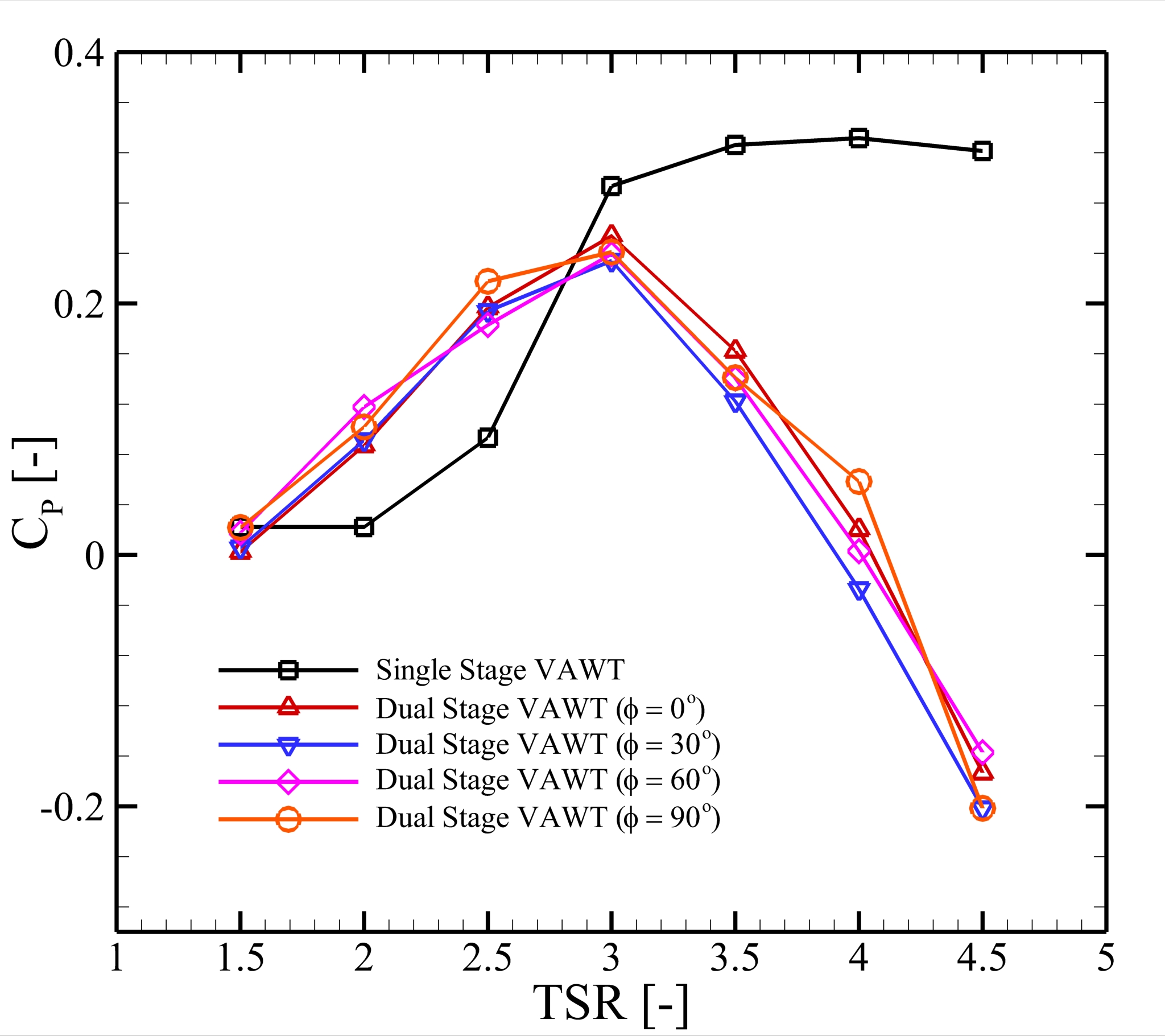}}
\subfigure[]{\includegraphics[scale=0.1]{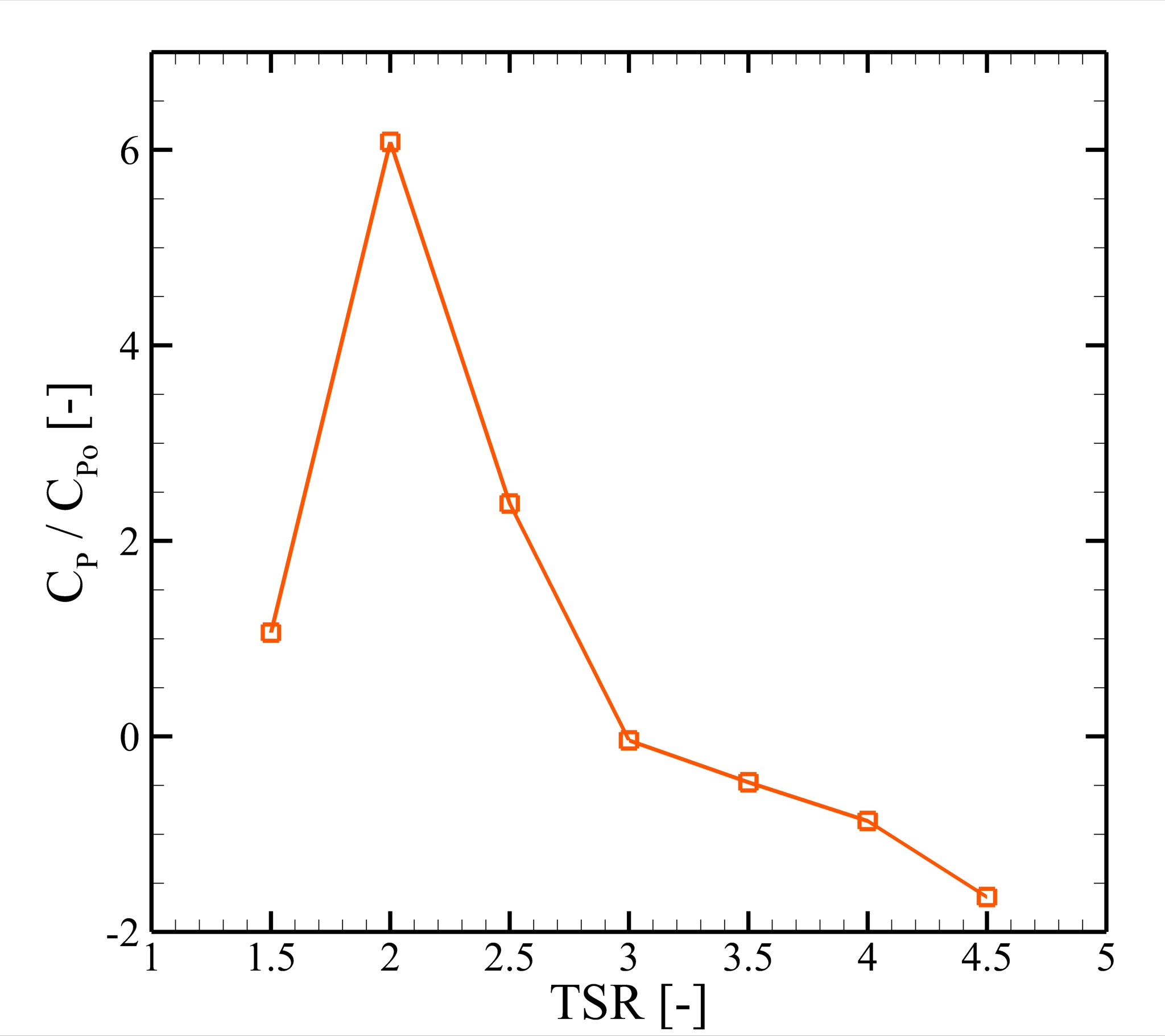}}
\caption{(a) Variations in $\overline{C_P}$ as a function of $\mbox{TSR}$ for the single-stage turbine and dual-stage VAWTs ($R_2/R_1=0.85$) for different values of $\phi$ (b) Ratio of power coefficients for a dual-stage turbine ($R_2/R_1=0.85$, $\phi=0^\circ$) and the single-stage VAWT}
\label{fig:Cp}
\end{figure}

We begin by looking at the aerodynamic performance parameters and flow analysis around dual-stage VAWTs in comparison to those for their single-stage counterparts. Here, $\lambda$ ranged from $1.50$ to $4.50$, whereas the geometric phase angle between the outer and inner rotors varied from $0^\circ$ to $90^\circ$. First, we presented a plot of $\overline{C_P}$ versus $\mbox{TSR}$ in Fig.~\ref{fig:Cp}a. It is apparent that the single-stage VAWT produced similar power for $\mbox{TSR}$ of $1.50$ and $2.0$. It enhanced with further increase in $\mbox{TSR}$ and reached its maxima for $\mbox{TSR}=4.0$. However, the difference in $\overline{C_P}$ was minimal for $ 3.50 < \mbox{TSR} < 4.50$. On the contrary, the dual-stage VAWTs performed significantly better than the single-stage turbine for $ 1.50 < \mbox{TSR} < 3.0$. Our data showed that introduction of the secondary rotor improved the power production by upto $400\%$ for this range of tip-speed ratios. These results also show that variations in $\phi$ did not substantially impact the energy harvesting capacity of dual-stage VAWTs and the trend for $ 0^\circ < \phi < 90^\circ$ remained the same. There were also negligible variations in $C_P$. However, the dual-stage VAWTs suffered from sharp decrements in $C_P$ for $\mbox{TSR} \ge 3.0$. This aspect of dual-stage turbines needed further investigations with variations in other important geometric parameters, including blades profiles and their aspect ratios. 

Upon increasing the ratio between the radii of primary (outer) and secondary (inner) rotors by bringing the inner blades closer to the outer ones, a substantial improvement in the performance of the dual-stage turbine was achieved for low tip-speed ratio as evident by the plot in Fig.~\ref{fig:Cp}b. Here, the ratio of $C_P$ to $C_{Po}$ was plotted, where $C_{P}$ belongs to a dual-stage turbine with $R_2/R_1=0.92$ and $\phi=0^\circ$ and $C_{Po}$ to a single-stage turbine. It showed improvements in power production by $600\%$ for $\mbox{TSR}=2.5$ and $225\%$ for $\mbox{TSR}=3.0$ by the dual-stage turbine. However, it was apparent from these results that $R_2/R_1 = 0.92$ did not suit the performance of the VAWT at higher $\mbox{TSRs}$.  

\begin{figure}[!ht]
\centering
{\includegraphics[scale=0.3]{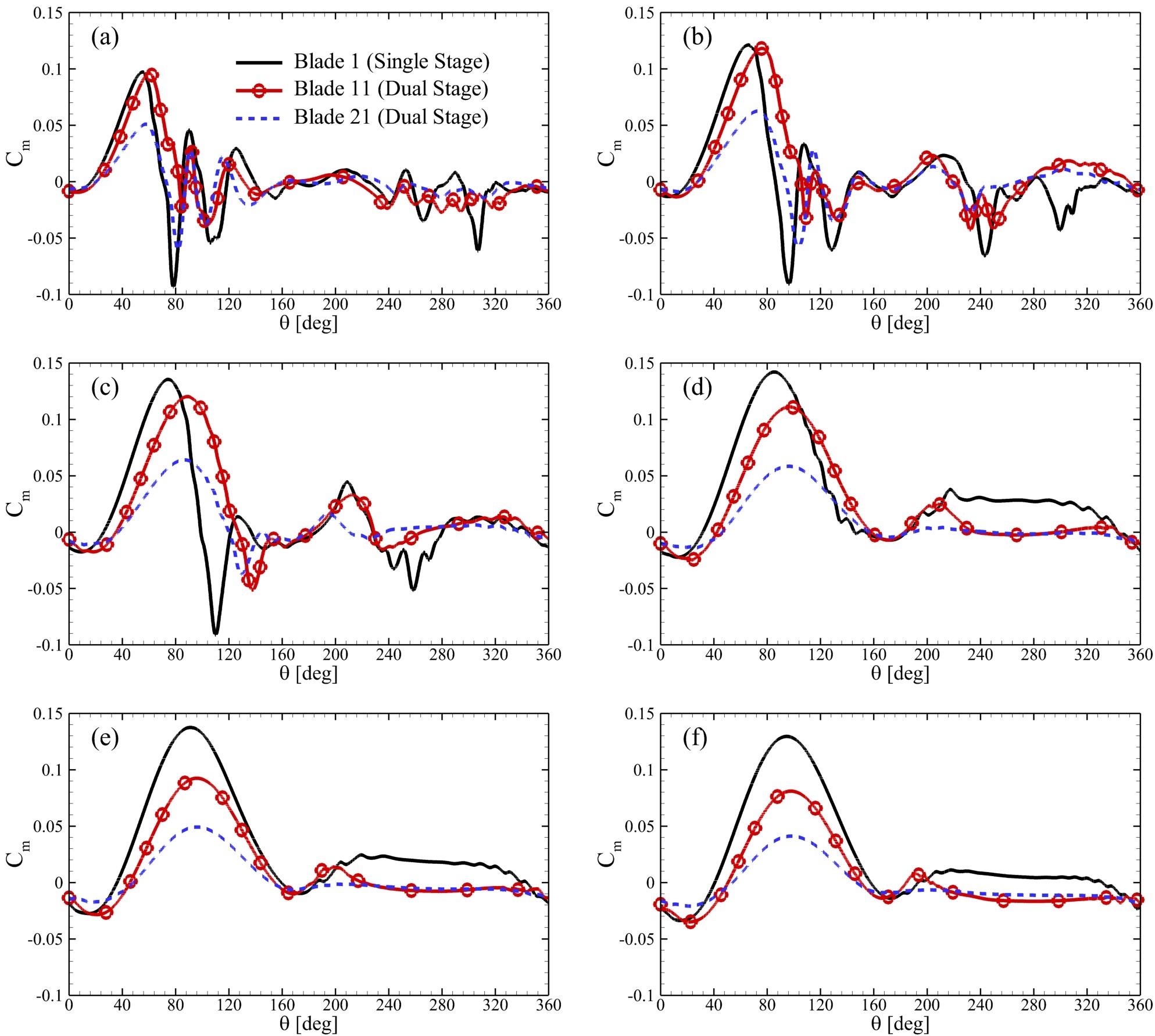}}
\caption{Variations in $C_m$ for single blades of the single-stage turbine, and outer and inner rotors of dual-stage VAWTs ($R_2/R_1$= 0.85) for $\phi=0^\circ$ and $\mbox{TSR}=$ (a) $1.50$, (b) $2.0$, (c) $2.50$, (d) $3.0$, (e) $3.50$, and (f) $4.0$}
\label{fig:Cm}
\end{figure}

In order to look deeper into the unsteady performance of these turbines, $C_m$ of single blades of the single-stage turbine were analyzed against primary and secondary rotors of dual-stage turbines in Fig.~\ref{fig:Cm}. It is important to highlight that the results were presented on the same scale so that the effect of $\mbox{TSR}$ could be demonstrated on the performance of these turbines. Here, the focus was on the performance parameters of the dual-stage VAWTs with $\phi=0^\circ$ due to insignificant effect of $\phi$ on their $\overline{C_P}$. The most important feature of this performance data was that blades in both stages of the dual-stage VAWT attained their maximum respective $C_m$ with specific delays for all $\mbox{TSR}$ in comparison to the one in the single-stage turbine. Here, Blade 11 and Blade 21 were used for single blades in the outer and inner rotors of the dual-stage VAWT. For $\mbox{TSR}=1.50$ in Fig.~\ref{fig:Cm}a, Blade 1 appeared to reach its maximum $C_m$ for $\theta=54^\circ$. However, Blades 11 and 21 attained this state at $\theta=60^\circ$. This implied that adding an auxiliary rotor helped delay the dynamic stall process, which was most likely induced by the ground effect provided by blades in the inner stage of the VAWT. The next important point involved avoiding large negative values of $C_m$ by Blades 11 and 21 at $\theta = 80^\circ$. It was evident that the blades of dual-stage VAWTs showed positive $C_m$ for a greater range of $\theta$, which demonstrated the possibilities of better self-starting capabilities of multi-stage turbines. Besides, Blades 11 and 21 experienced significant reduction in magnitudes of their respective $C_m$ for $TSR > 3.0$ as is also shown in their $\overline{C_P}$ in Fig.~\ref{fig:Cp}. A plausible reason for better performance of the single-stage VAWT for higher $\mbox{TSR}$ can be that its blades stopped producing negative $C_m$ for $\theta > 100^\circ$, while there were negative values observed for a single stage system at similar conditions.        

\begin{figure}[!ht]
\centering
{\includegraphics[scale=0.3]{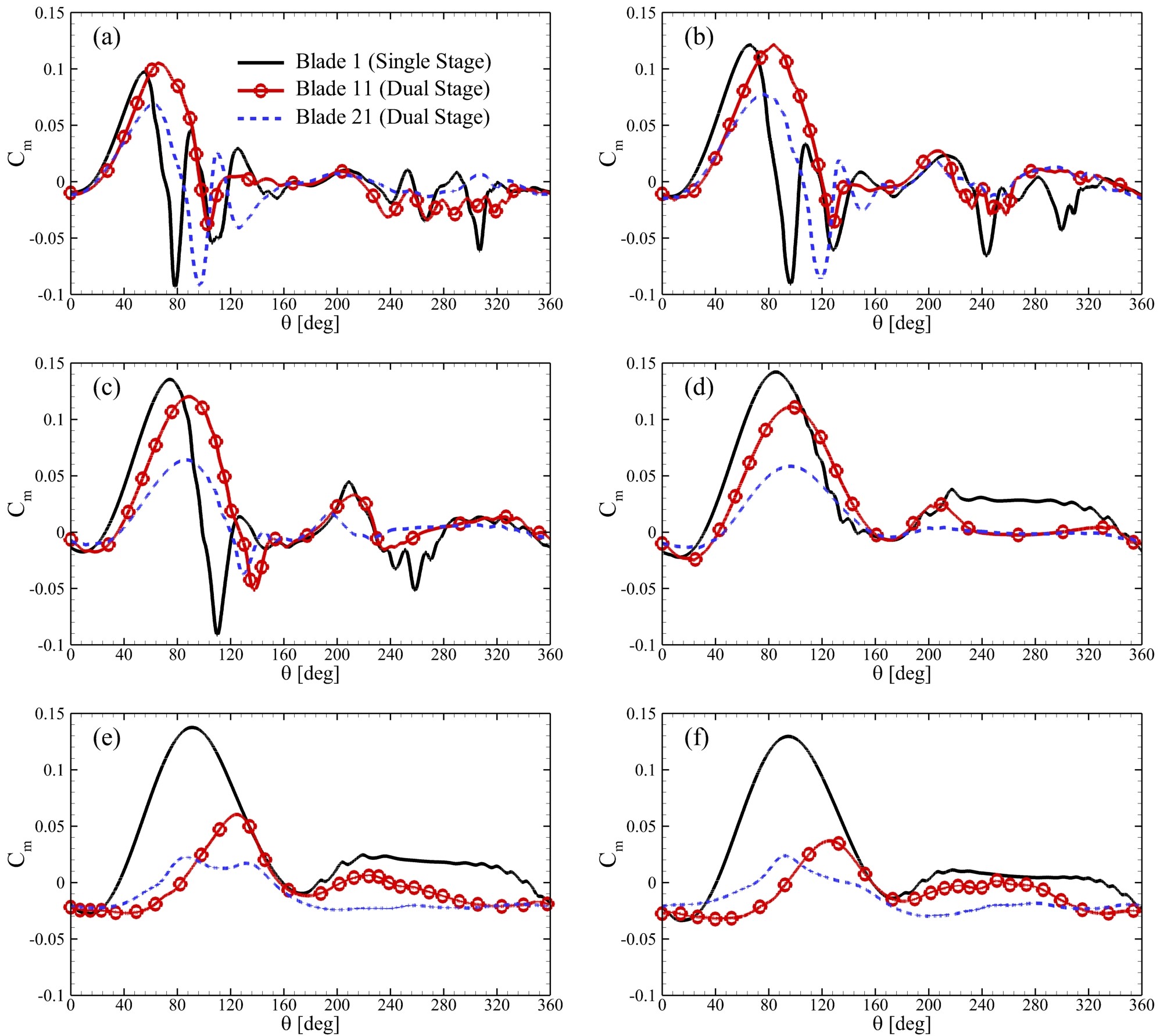}}
\caption{Variations in $C_m$ for single blades of the single-stage turbine, and outer and inner rotors of dual-stage VAWTs ($R_2/R_1$= 0.92) for $\phi=0^\circ$ and $\mbox{TSR}=$ (a) $1.50$, (b) $2.0$, (c) $2.50$, (d) $3.0$, (e) $3.50$, and (f) $4.0$}
\label{fig:Cm2}
\end{figure}

Moreover, Fig.~\ref{fig:Cm2} presents temporal profiles of moment coefficients of single-stage and dual-stage ($R_2/R_1=0.92$ and $\phi=0^\circ$) VAWTs. Here, Blade 11 exhibited a higher $C_m$ than Blade 1 at a greater azimuthal angle (see Fig.~\ref{fig:Cm2}a). For this blade of the primary rotor, negative value of $C_m$ was also significantly reduced. Despite these improvements, Blade 21 suffered from a negative $C_m$ with magnitudes that almost matched the order of Blade 1, contrary to what we observed for the inner rotor blade for the VAWT with $R_2/R_1=0.85$ in Fig.~\ref{fig:Cm}a. It is also important to highlight that $C_m$ of Blade 21 was lower than that of Blade 1 mostly for $\theta > 200^\circ$ at $\mbox{TSR}=1.50$. However, it showed improvements in $C_m$ for these azimuthal positions at $\mbox{TSR}=2.0$ and $2.50$ in Figs.~\ref{fig:Cm2}b and \ref{fig:Cm2}b, respectively. The blades of the dual-stage turbine started showing signs that implied their reduced performance from $\mbox{TSR}=3.00$ in Fig.~\ref{fig:Cm2}d and both rotors individually produced less power compared to the single-stage version for almost the whole range of $\theta$. It was even deteriorated for $\mbox{TSR} \ge 3.50$.            

\begin{figure}[!ht]
\centering
{\includegraphics[scale=0.12]{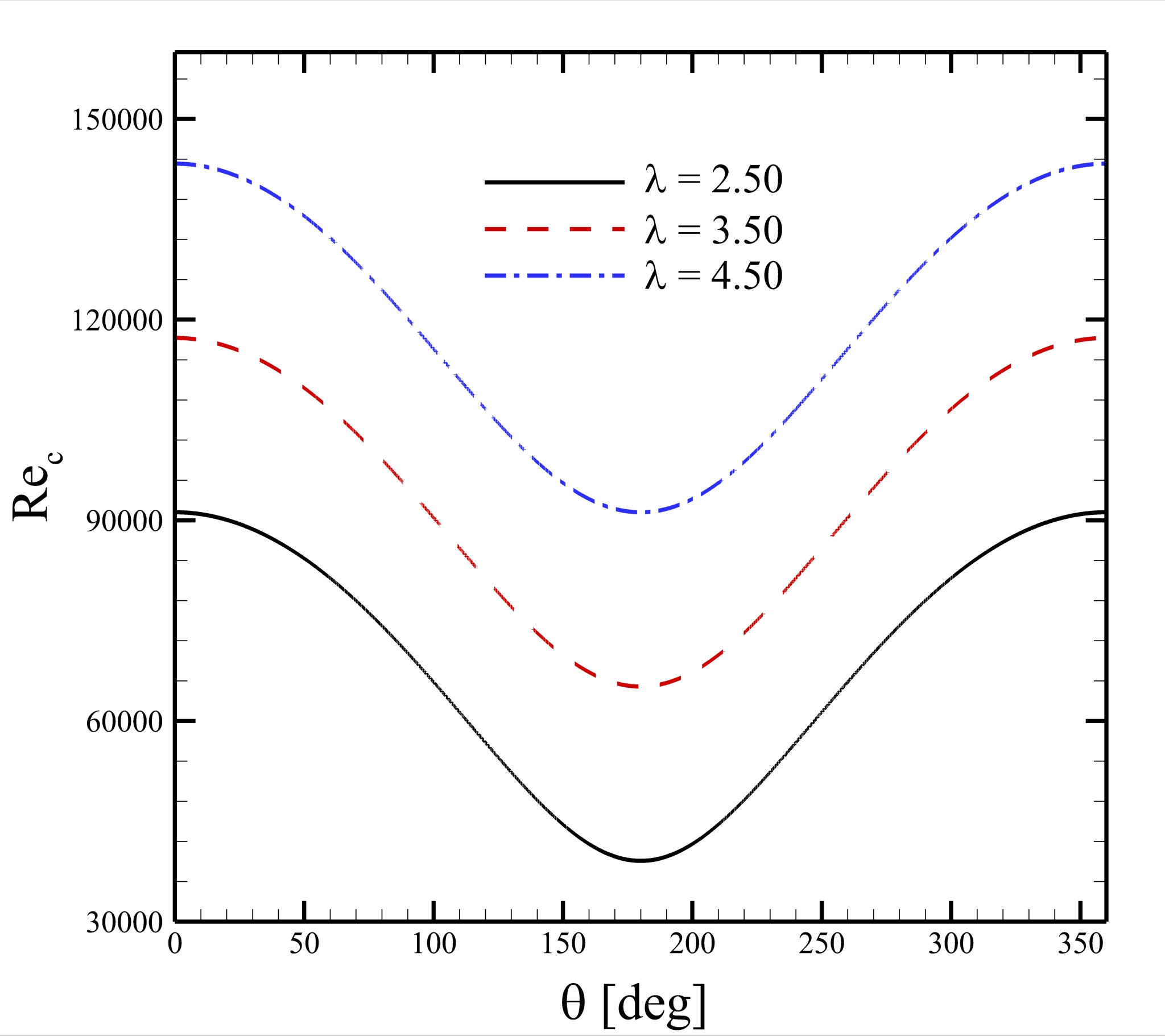}}
\caption{Variations in chord-based Reynolds numbers with respect to azimuthal positions of a blade at different tip-speed ratios}
\label{fig:Re}
\end{figure}

This behavior of single- and double-stage VAWTs can be further elucidated by examining their production of lift and drag with respect to effective angles-of-attack of the blades. For this purpose, variations in Reynolds numbers throughout their rotational cycles were shown for different $\mbox{TSRs}$ in Fig.~\ref{fig:Re}. Because the effective velocity of blades; a function of their azimuthal positions, was employed as the velocity scale for the computation of $\mbox{Re}$, its variation follows a cosine function when plotted against $\theta$. It was evident that for each $\mbox{TSR}$, the difference in the maximum and minimum $\mbox{Re}$ was around $40,000$. It showed extreme unsteadiness experienced by these blades during their rotation. To avoid this complexity, an average $\mbox{Re}$ was chosen for further comparative analysis for aerodynamics of rotating turbines and static airfoils.  

\begin{figure}[!ht]
\centering
{\includegraphics[scale=0.23]{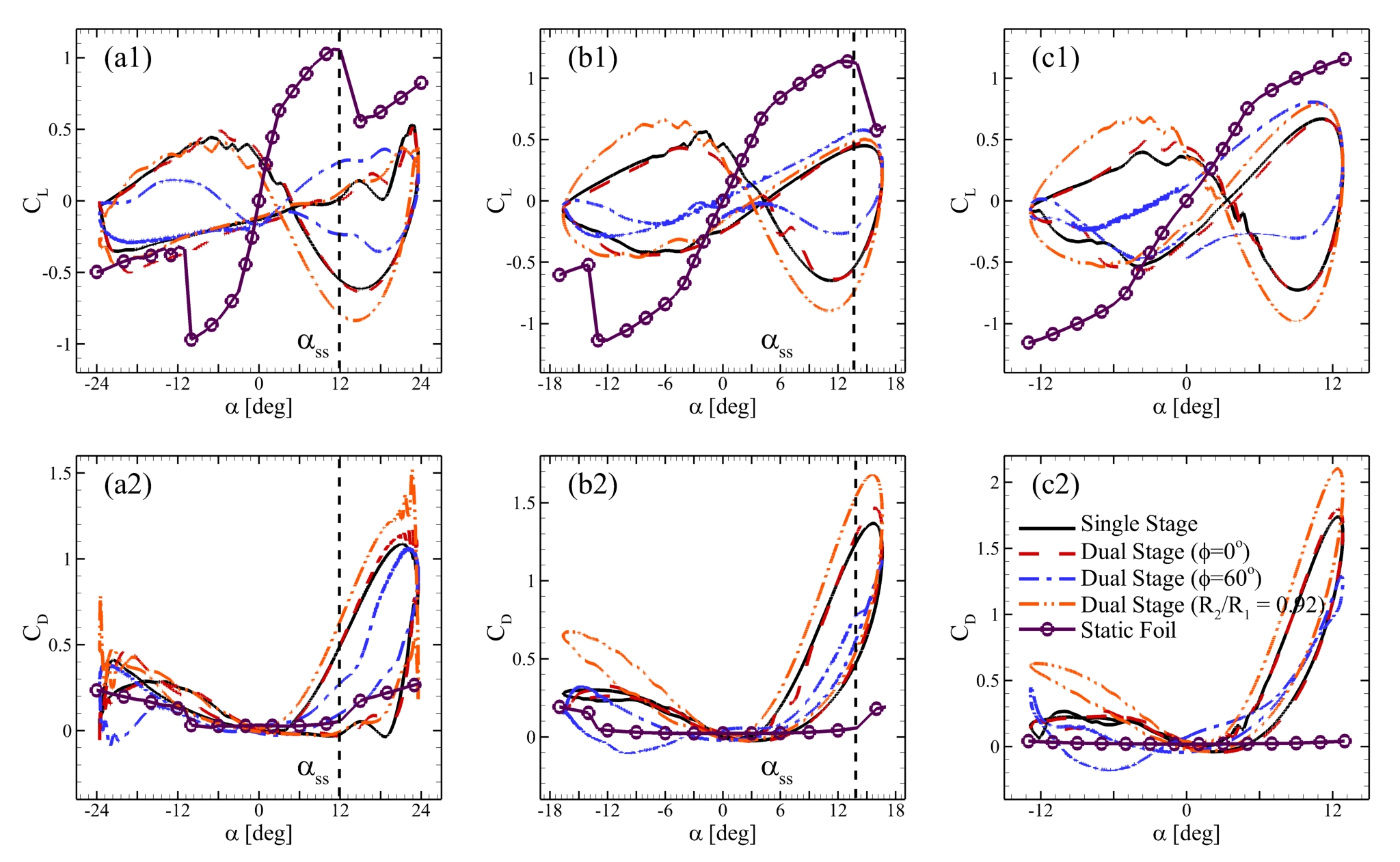}}
\caption{Variations of dynamic loads on a single blade with respect to $\alpha$ where (a1) and (a2) correspond to $\mbox{TSR=2.50}$, (b1) and (b2) are for $\mbox{TSR}=3.50$, and (c1) and (c2) present data for $\mbox{TSR}=4.50$}
\label{fig:ClCd}
\end{figure}

Figure~\ref{fig:ClCd} presents temporal variations in lift and drag coefficients of single blades of VAWTs with respect to $\alpha$, denoted as $C_L$ and $C_D$, respectively. Here, the data for Blade 11 of dual-stage turbines was plotted for simplicity and brevity. To illustrate the effect of $\phi$ of the unsteady aerodynamics of these rotating blades, $C_L$ and $C_D$ for the VAWT with $\phi=60^\circ$ were included as well. In Fig.~\ref{fig:ClCd}a1, the rotating blades experienced less lift for $\alpha > 2^\circ$. A static airfoil experiences stall at $\alpha_{ss}=12^\circ$, and the blades in the single-stage VAWT and dual-stage turbines attained their maximum $C_L$ for $\alpha > \alpha_{ss}$ and the entire range of $\phi$ was considered here. The data for static foils was computed using XFOIL \cite{drela1989xfoil}. These results also present that the blades of the single-stage turbine and dual-stage VAWT with $\phi=0^\circ$ showed almost similar hysteresis. However, the dual-stage turbine blade with $\phi=60^\circ$ exhibited larger hysteresis. These observations also held for $\mbox{TSR}=3.50$ in Fig.~\ref{fig:ClCd}b1. For $\mbox{TSR}=4.50$ in Fig.~\ref{fig:ClCd}c1, the blades did not undergo $\alpha > \alpha_{ss}$ at any stage of the rotational cycle. It appears that flow remained attached to the blades for all $\alpha$, but the pressure difference for their outer and inner surfaces was less than that required to have $C_L$ at any $\alpha$. Moreover, the blades for these VAWTs experienced $C_D$ greater than that of a static airfoil for most ranges of $\alpha$. Moreover, dual-stage turbine blades showed larger hysteresis for both $\phi$ compared to that of the single-stage VAWT. 

\begin{figure}
\centering
{\subfigure{\includegraphics[width=0.4\textwidth]{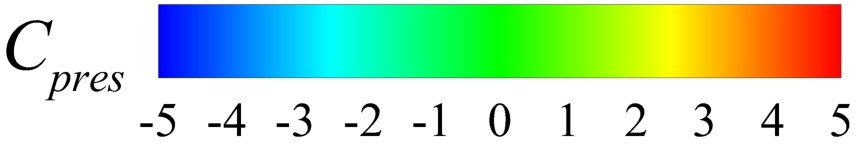}}}\\
\subfigure{\includegraphics[scale=0.18]{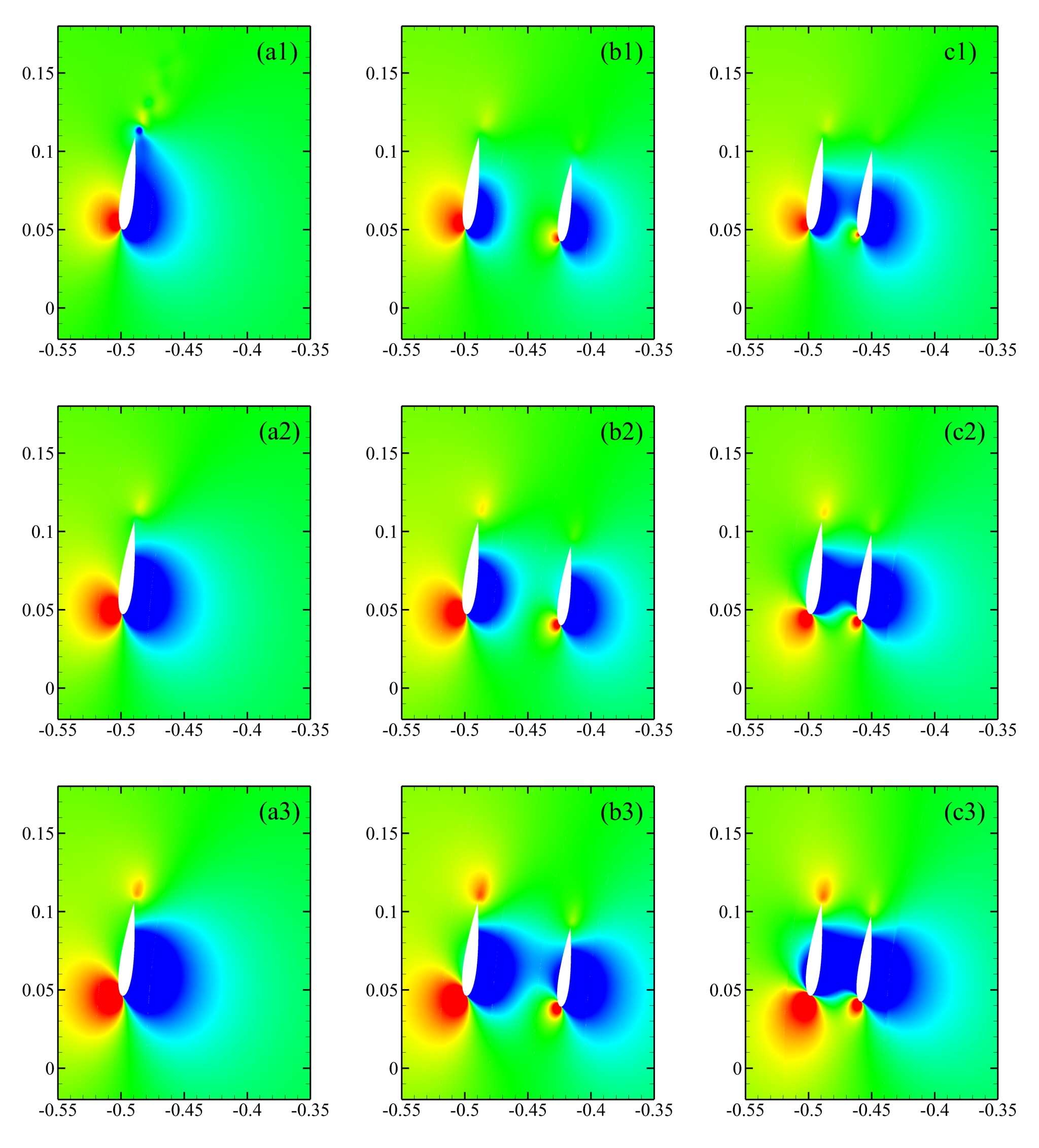}}
\caption{Contours of $C_{pres}$ around the blades of a single-stage turbine (left column), dual-stage turbine with $\phi=0^\circ$ \& $R_2/R_1=0.85$ (middle column), and dual-stage turbine with $\phi=0^\circ$ \& $R_2/R_1=0.92$ (right column) positioned at $\theta=81^\circ$, whereas the top, middle, and bottom rows correspond to $\mbox{TSR}=2.50$, $3.50$, and $4.50$, respectively}
\label{fig:Cpress}
\end{figure}

These trends in dynamic loads were then related with pressure variations around the blades through contours of the pressure coefficient ($C_{pres}$) in Fig.~\ref{fig:Cpress}. For $\mbox{TSR}=2.50$ (Fig.~\ref{fig:Cpress}a1, \ref{fig:Cpress}b1 and \ref{fig:Cpress}c1), there appeared to be a large difference in the distributions of $C_p$ around the blades for single- and double-stage VAWTs. A low-pressure region expanded over the whole inner surface of the blade of the single-stage turbine, which was contracted due to the presence of the inner rotor in dual-stage turbines. These pressure distributions also explained the lower power production by the inner rotors. Outer surfaces of its blades did not have significantly higher pressure regions formed over them, which was more prominent for the dual-stage turbine with $R_2/R_1=0.92$. At higher $\mbox{TSRs}$, both high- and low-pressure regions expanded for the blade of a single-stage turbine (see Figs.\ref{fig:Cpress}a2 and \ref{fig:Cpress}a3). It contributed to produce more lift by the blade that consequently improved its power production. Similar observations were made for blades of the dual-stage turbines as well in Figs.~\ref{fig:Cpress}b2 and \ref{fig:Cpress}b3, and Figs.\ref{fig:Cpress}c2 and \ref{fig:Cpress}c3. In Fig.~\ref{fig:Cpress}c3, the formation of a low pressure region expanded over almost half of the outer surface of the blade in the primary rotor. Perhaps, such phenomena were responsible for the sudden deterioration of its aerodynamic performance for higher $\mbox{TSRs}$ shown in Fig.~\ref{fig:Cm2}.         

\begin{figure}
\centering
{\subfigure{\includegraphics[width=0.4\textwidth]{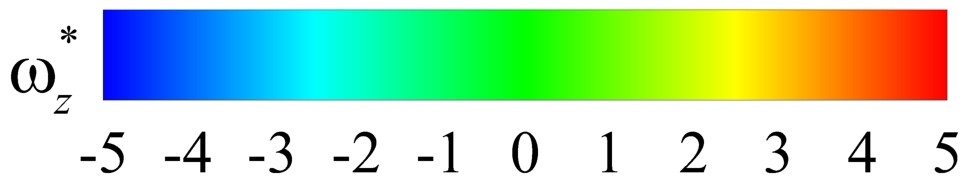}}}\\
\subfigure{\includegraphics[scale=0.2]{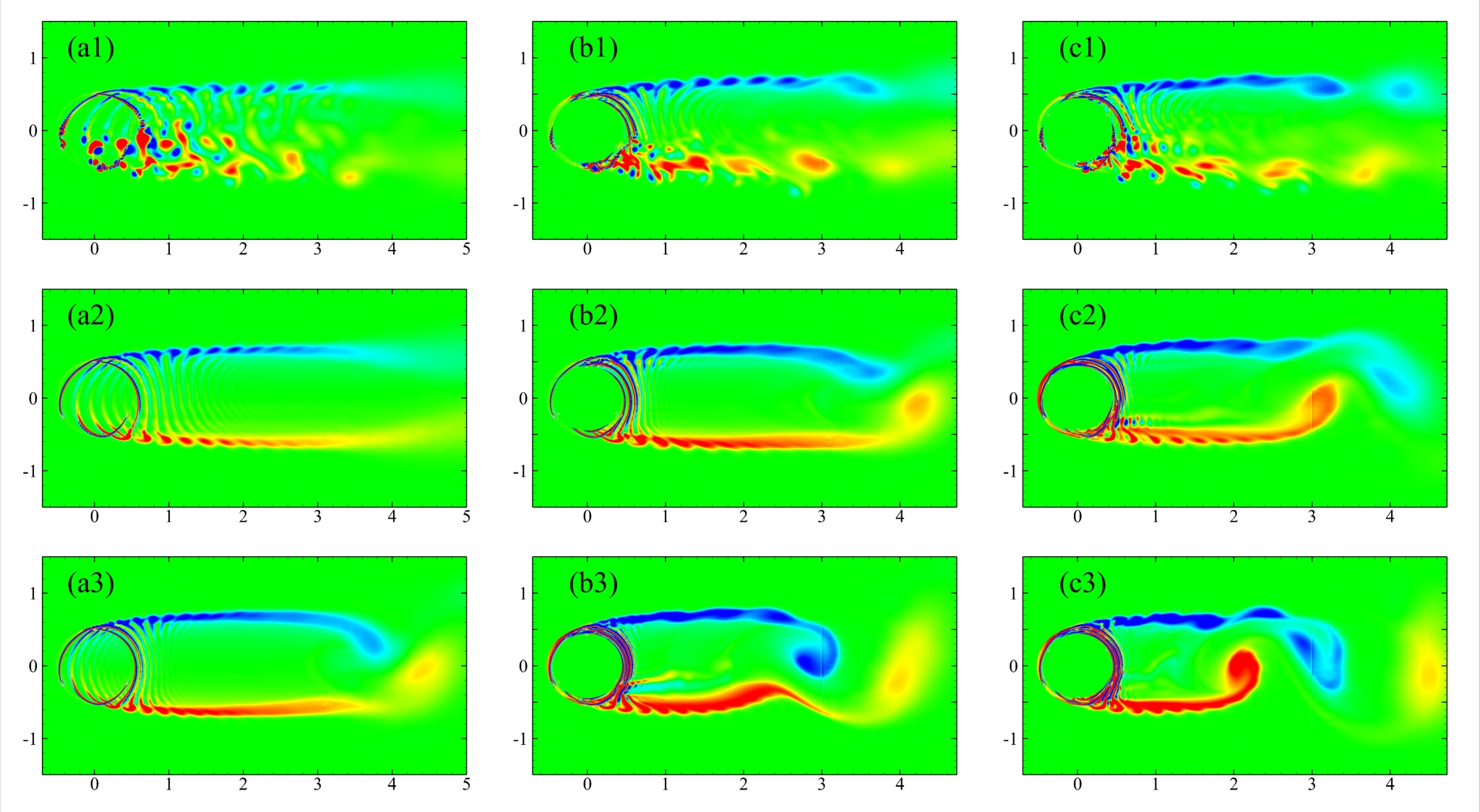}}
\caption{Contours of nodimensional vorticity (${\omega_z}^* = {\omega}D/U_\infty$) around the blades of a single-stage turbine (left column), dual-stage turbine with $\phi=0^\circ$ \& $R_2/R_1=0.85$ (middle column), and dual-stage turbine with $\phi=0^\circ$ \& $R_2/R_1=0.92$ (right column) positioned at $\theta=81^\circ$, whereas the top, middle, and bottom rows correspond to $\mbox{TSR}=2.50$, $3.50$, and $4.50$, respectively}
\label{fig:wake}
\end{figure}

Finally, the details of overall wake formations behind single- and dual-stage turbines are shown in Fig.~\ref{fig:wake}. Figures~\ref{fig:wake}a1, \ref{fig:wake}b1, and \ref{fig:wake}c1 show vorticity contours of the flow around a single-stage turbine, and double-stage VAWTs with $R_2/R_1=0.85$ and $R_2/R_1=0.92$, respectively. There existed discrete vortices shed by the foil at $\theta=0^\circ$ for the single-stage VAWT. Near the opposite end of the turbine, complex vortical structures were present, which indicated extensive vortex-blade interactions. Such phenomena were more noticeable for $90^\circ < \theta < 270^\circ$. For dual-stage turbines, the vortices were more interconnected with each other with enhanced vorticity levels in the respective wakes due to the additional contribution by the inner-stage rotors. The upper vortex array was thickened for the dual-stage VAWT with the greater ratio of radii. This observation hinted at more constructive interactions \cite{khalid2021anguilliform} between the vortices shed by primary and secondary blades in this case. This may also be the primary reason for higher drag production at this $\mbox{TSR}$ possibly resulting in more torque experienced by the turbine (see Fig.~\ref{fig:Cm2}). At higher $\mbox{TSRs}$, the wake of turbines resembled those of circular cylinders of large diameters, which are shown in the plots of middle and bottom rows of Fig.~\ref{fig:wake}. However, the shedding of big vortices, with length scales of almost the order of the radii of turbines, was more prominent for dual-stage VAWTs. Under these conditions, the shear layer starts rolling at a distance of $3.50D$ and $3D$ from their centers in Figs.~\ref{fig:wake}b2 and \ref{fig:wake}c2, respectively, for turbines rotating with $\mbox{TSR}=3.50$. This formation of big vortices not only enhanced drag production for these conditions, but also it would affect the performance of any turbines installed downstream. For even higher $\mbox{TSRs}$, such as the one in Fig.~\ref{fig:wake}a3, the single-stage turbine also showed the signs of this vortex shedding at a distance of $4D$ from its center, which may be related to decrements in the performance of the single-stage turbines, as shown in Fig.\ref{fig:Cp}a. For dual-stage VAWTs in Figs.~\ref{fig:wake}b3 and \ref{fig:wake}c3, this shedding process was more intense and started in the regions closer to the turbines. This also contributed to the reduced performance of these turbines at higher tip-speed ratios.      

\section{Conclusions}
\label{sec:concl}
The flow around single- and dual-stage vertical-axis wind turbines were numerically examined for $\mbox{TSR}$ ranging from $1.50$ to $4.50$. It was determined that dual-stage VAWTs outperformed their single-stage counterparts by enhancing power production manifolds at $ 1.50 < \mbox{TSRs} < 3.0$. However, their performance substantially degraded for higher $\mbox{TSR}s$. The geometric phase angle between the blades in two rotors of dual-stage turbines did not significantly impact their performance in a time-averaged sense. However, bringing these rotors closer had a greater impact on their power production. This change enhanced their $C_P$ from $400\%$ to $600\%$ at $\mbox{TSR}=2.0$. The analysis of underlying flow mechanisms associated with trends of performance parameters of these turbines revealed contributing factors for degraded power generation of dual-stage turbines at higher tip-speed ratios: (i) the formation of low-pressure regions over the outer surfaces of the blades of primary rotors, which adversely affects their lift production; (ii) the rolling of shear layers in the wake of these turbines, and forming and shedding of large vortices. These coherent structures with length scales of the order of radii of these VAWTs generated greater drag. Further investigations are required to examine the impact of other important parameters to better understand the performance and self-starting capabilities of dual-stage turbines.  


\section*{Declaration of competing interest}
The authors declare that they have no known competing financial interests or personal relationships that could have appeared to influence the work reported in this paper.

\section*{Acknowledgment}
This study has received support from Future Energy Systems (through Canada First Research Excellence Fund) with project number T14-Q01.

\bibliographystyle{model3-num-names}
\bibliography{MyBibFile_Complete}
\end{document}